\documentclass[runningheads]{llncs}
\usepackage{amssymb}
\usepackage{graphicx}
\usepackage{cite}
\usepackage{color}
\usepackage{textcomp}
\usepackage{url}
\usepackage{caption}
\usepackage{siunitx}

\usepackage{academicons}
\usepackage{hyperref}
\usepackage{xcolor}
\newcommand{\orcid}[1]{\href{https://orcid.org/#1}{\includegraphics[scale=0.08]{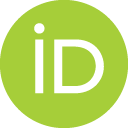}}}

\usepackage{url}
\urldef{\mailsa}\path|{alfred.hofmann, ursula.barth, ingrid.haas, frank.holzwarth,|
\urldef{\mailsb}\path|anna.kramer, leonie.kunz, christine.reiss, nicole.sator,|
\urldef{\mailsc}\path|erika.siebert-cole, peter.strasser, lncs}@springer.com|    
\newcommand{\keywords}[1]{\par\addvspace\baselineskip
\noindent\keywordname\enspace\ignorespaces#1}

\hypersetup{colorlinks=false}

\begin{document}

\mainmatter  

\title{Gas Adsorption on Graphtriyne Membrane: Impact of the Induction Contribution on the Computational Cost}

\titlerunning{Gas Adsorption on Graphtriyne: the Computational Cost of Induction}
\author{Emília Valença Ferreira de Aragão\inst{1,2}\orcid{0000-0002-8067-0914} 
\and Noelia Faginas-Lago\inst{1}\orcid{0000-0002-4056-3364} 
\and Yusuf Bramastya Apriliyanto\inst{3}\orcid{0000-0003-0683-8456} 
\and Andrea Lombardi\inst{1}\orcid{0000-0002-7875-2697} \\\
}
\authorrunning{E.V.F. Aragão et al.}

\institute{Dipartimento di Chimica, Biologia e Biotecnologie,\\ Universit\`{a} degli Studi di Perugia, 06123 Perugia, Italy\\ 
\email{emilia.dearagao@studenti.unipg.it}\\
\email{\{noelia.faginaslago,andrea.lombardi\}@unipg.it}\\
\and
Master-up srl, Via Sicilia 41, 06128 Perugia, Italy\\
\email{emilia.dearagao@master-up.it}
\and
Department of Chemistry, Bogor Agricultural University, Jl Tanjung Kampus IPB Dramaga, 16680, Bogor, Indonesia\\
}

\toctitle{Lecture Notes in Computer Science}
\tocauthor{Authors' Instructions}
\maketitle

\begin{abstract}

Graphynes are a family of porous carbon allotropes that are viewed as ideal 2D nanofilters.
In this present work, the authors have modified the Improved Lennard Jones (ILJ) semi-empirical potential used in the previous works by adding the induction term (iind) to define the full interaction.
The evaluation of the computational cost was done comparing ILJ vs ILJ-iind and analyzing the adsorption of 1 gas (CO$_{2}$) and a small mixture of gases containing CO$_{2}$, N$_{2}$ and H$_{2}$O.
The computational time of the different calculations is compared and possible improvements of the potential models are discussed.

\end{abstract}

\section{Introduction}

Recent reports have shown that the concentration of CO$_{2}$ in the atmosphere has risen a lot the last few decades~\cite{epa}.
This trend is seen as a consequence of large-scale human activity, whether it involves energy production or manufacturing materials (cement, iron, steel, etc)~\cite{smit16:9}.
The excess of CO$_{2}$ in the atmosphere causes many problems, such as the more frequent apparition of toxic blue green algae in lakes during hot seasons and the rising of global temperatures~\cite{wri}.
This is why it is urgent to investigate strategies that can be implemented in order to, if not lower the CO$_{2}$ in the atmosphere, at least change the evolution trend and keep the CO$_{2}$ concentration at the current level.
There are two main ways to do it: either capturing CO$_{2}$ in the open air or in the place where it is produced. 
The second approach is generally seen as more efficient and energetically cheaper than the first, but there are some constraints: the flue gas is a mixture of water, carbon dioxide, oxygen and nitrogen molecules~\cite{SONG2004315}.
One way to selectively capture CO$_{2}$ in flue gas is through the adsorption using porous materials~\cite{huck14:4132,bui18:1062,li13:1538,celiberto2016atomic}. 
The advantage of this method is that it is relatively cheaper and simpler to implement on existing power plants~\cite{smit16:9}.
In the last few years, a range of porous materials have been evaluated in their ability to selectively capture CO$_{2}$: a) nanoporous carbons~\cite{srinivas14:335,ganesan15:21,ghosh16:14739}, b) zeolites and zeolitic imidazolate frameworks (ZIFs)~\cite{kim12:18940,liu10:8515}, c) metal-organic frameworks (MOFs)~\cite{lin13:4410}, d) porous polymer networks (PPNs) or covalent organic frameworks/polymers (COFs/COPs)~\cite{schrier12:3745,xiang15:13301}, and e) a slurry made of solid adsorbents in a liquid absorbent~\cite{liu14:5147}.
In particular, carbon-based membranes have desirable physicochemical properties (e.g. hydrophobic, chemically inert and thermally stable) and are economically suitable and viable for carbon capture and sequestration (CCS)~\cite{dubay12:4556,bartolomei15:1076}.
In contrast, MOFs show permeability and good selectivity, but they are not resistant in the presence of water vapor nor in high temperature, which are usually the conditions of CO$_{2}$ combustion.
 What makes both of these classes effective for gas separation is their permeability and their selectivity.
In addition, the thinner a membrane is, the more it is permeable, which makes single-layer membranes interesting objects of study in this context~\cite{du11:23261}.

In CCS the range of options of applicable materials is vast, so it is impossible to do the synthesis, the characterization and the evaluation of the ability to selectively capture CO$_{2}$ for all of the candidates~\cite{lombardi12:387,lombardi13:11430,falcinelli13:69}.
This is why a preliminary investigation with computational modelling and simulation is crucial for narrowing the selection of molecules. 
Molecular Dynamics (MD) simulations are a theoretical chemistry method for analyzing the movement of atoms and molecules using potential functions.
It can be employed to investigate the structural rearrangement of pure solvents, mixed solutions and combustion processes.~\cite{pallottelli2010distributed,Lago2006,lagana2008thermal,rampino2012extension,Lagana2003,FaginasLago2013,Lombardi2014}.
In MD, a set of potential functions is called a force field. 
Currently, a number of options for force fields are available in MD software, such as UFF~\cite{rappe92:10024} and AMBER~\cite{pearlman95:1}.
However, both these force fields are limited in use: when studying a particular system, those force fields do not always have appropriate parameters to describe it since they are too generic.
Therefore, the researcher interested in a particular system has to develop or modify parts of the force field, choosing better potential energy functions.
The choice of the modifications must be based on experimental and theoretical data available from the literature or prior quantum chemical computations.
While the parameterization of force fields is not a simple task, it is crucial for describing a system correctly.
Recently, a number of force fields have been developed for evaluating the adsorption of gas molecules on different porous materials like zeolites~\cite{lim18:10892}, MOFs~\cite{boyd17:357,lin14:1477}, graphene and its derivatives~\cite{vekeman18:25518,Lombardi2015,faginas14:2226} and other polymeric materials~\cite{dubay12:4556}.
These force fields are used to describe molecular interactions between the gases and the porous layer in a quantitative manner, aiming to give predictions of the adsorption dynamics and transport properties of the gases.

The authors of this paper have been recently involved in the study of $\gamma$-graphynes using MD tools, in particular the development of force fields related to gas adsorption in that class of carbon allotropes.~\cite{Faginas2008,yusuf18:16195,Faginas-Lago2016,yeamin14:54447,MOLSIM2019}.
$\gamma$-graphynes are atomic monolayers where the carbon atoms are arranged in a way that two adjacent hexagons are connected through C-C triple bonds. 
As nanoporous materials, $\gamma$-graphynes are interesting candidates for CCS.
It has been reported in the literature that the pores are uniformly distributed and adjustable~\cite{james18:22998} and that they are not prone to form aggregates due to low dispersion forces.
In this work, simulations were performed involving a form of $\gamma$-graphyne called graphtriyne and a mixture of CO$_{2}$/N$_{2}$/H$_{2}$O in an attempt to reproduce the chemical environment of flue gases.
A challenge the authors have been facing is the considerable increase of the computational time by many times because of a choice made in the potential function to represent the interaction between the gases. 
In this report, a discussion is hold on how this problem is being confronted.

In the next section of the present work, methods and construction of the present potential energy function are outlined. In Sec.~3 a discussion about the improvements of the code is hold and in Sec.~4 the paper brings up concluding remarks.

\section{Methods}

\begin{figure}
\centering
\includegraphics[scale=0.36]{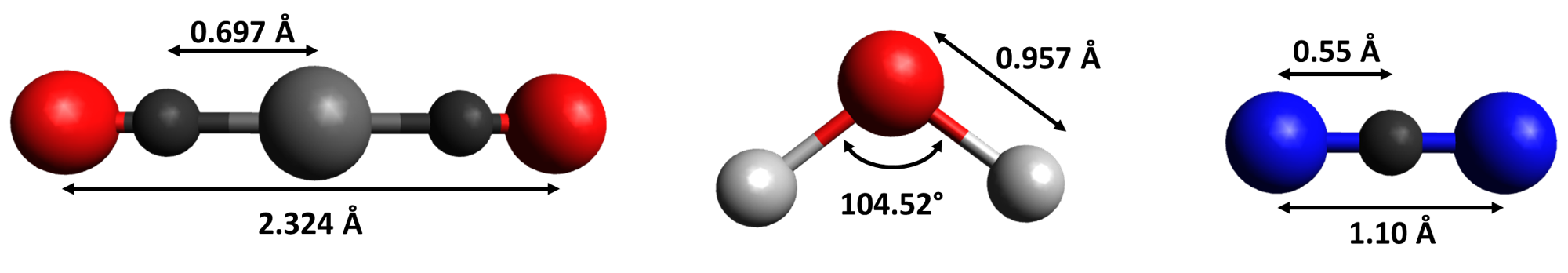}
\caption{Structural details of the model representation of carbon dioxide, water and nitrogen molecules. Bond lengths are shown in \si{\angstrom} and water's bond angle is displayed in degrees.}
\label {fig1}
\end{figure}

\begin{figure}
\centering
\includegraphics[scale=0.40]{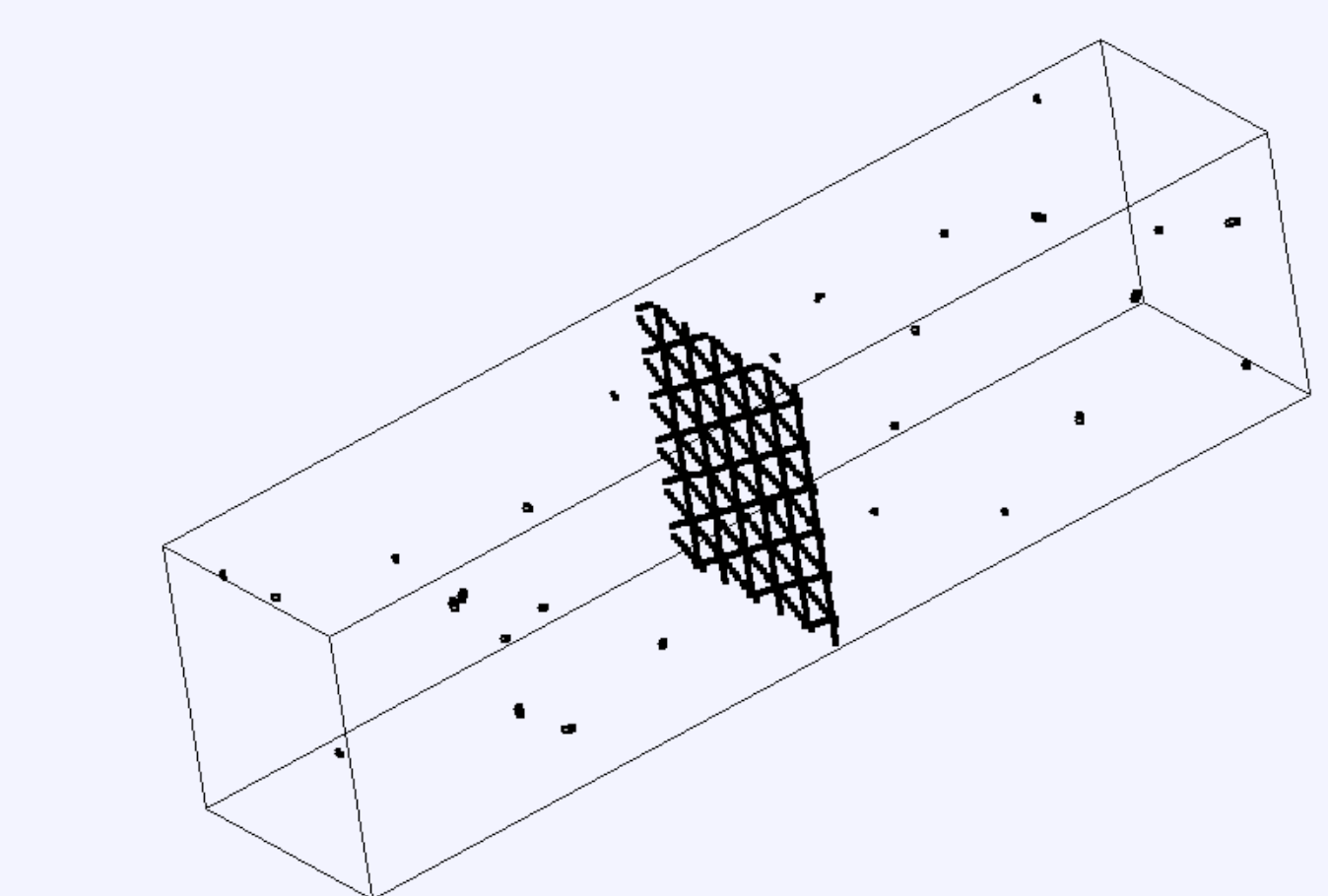}
\caption{A screenshot of a simulation box with gaseous mixture of CO$_{2}$, N$_{2}$ and H$_{2}$O. The layer in the middle of the box is the graphtriyne membrane.}
\label {fig2}
\end{figure}

The MD simulations were performed in simulation boxes with dimensions 72.210 \AA~x 62.523 \AA~x 280.0 \AA. 
Inside each box, a graphtriyne membrane with dimensions 72.210 \AA~x 62.523 \AA~ was placed. The membrane structure was taken from Ref.~\cite{bartolomei16:27996}, and had been previously optimized through periodic DFT calculations.
Simulations were performed uniquely at the temperature of 333 K.
Part of the simulations were performed with only CO$_{2}$ molecules, while the other part involved a CO$_{2}$/N$_{2}$/H$_{2}$O gaseous mixture. This mixture was composed of an equal number of moles of all molecules.
For the nitrogen and the carbon dioxide molecules, models taking into account the quadrupole moment were employed. Those models were a three-charge-site N$_{2}$ model~\cite{lombardi16:1463} and a five-charge-site CO$_{2}$ model~\cite{bartolomei12:1806}.
As for the water molecule, a model taken from Ref.~\cite{alberti09:50} was used. In this model, the charge is distributed in a way that corresponds to the dipole moment of water in the gas phase (1.85 D)~\cite{alberti13:6991}.
All the details of the geometries of those molecules are shown in Figure~\ref{fig1}. 
For the MD simulation, the membranes were set as a frozen framework while the gas molecules were set as rigid bodies.
The gas molecules were randomly distributed with equal amount into each region of the box (Figure~\ref{fig2}).

In a MD simulation, defining correctly the intermolecular forces at play is important for obtaining accurate results. 
In this system, the intermolecular forces of interest are those between gas molecules and between the gas molecules and the graphtriyne membrane.
The intermolecular interaction energy is decomposed in terms of molecule-molecule pair contribution, which are electrostatic and non-electrostatic contributions.
The non-electrostatic contributions are measured by taking into consideration the strength of induced dipoles and the average molecular sizes.
This can be done by assigning a value of polarizability to the both interacting centers, as shown in Figure~\ref{fig1}.
Here, the intermolecular forces were expressed using the Improved Lennard-Jones (ILJ) potential~\cite{pirani08:5489,Lombardi2013,Lago2014,FaginasLago2016,Faginas-Lago2015,faginas13:1,faginas15:192,lombardi16:246,alberti2012:3094}.

\begin{equation}V_{ILJ}(r) =
\varepsilon \left[\frac{m}{n(r)-m}\left (
\frac{r_{0}}{r} \right)^{n(r)} -
\frac{n(r)}{n(r)-m}\left(
\frac{r_{0}}{r}\right)^m \right]\label{eq:1}
\end{equation}

In Eq.~\ref{eq:1}, $\varepsilon$, $r_{0}$ and $m$ are parameters specific to the molecular pair involved, and $r$ is the distance between the two interacting centers of the same molecular pair.
In particular, m assumes the value of 6 for neutral-neutral pairs, 4 for ion-neutral pair and 1 for ion-ion pairs.
The first term of the Eq.~\ref{eq:1} represents the dependence of the repulsion in function of $r$, while the second term is the dependence of the long-range attraction in function of $r$.
To modulate the decline of the repulsion and the strength of the attraction, the $n(r)$ term is employed (Eq.~\ref{eq:2}).

\begin{equation}
n(r) = \beta +4.0\left(\frac{r}{r_{0}}\right)^2\label{eq:2}
\end{equation}
\noindent

In Eq.~\ref{eq:2}, $\beta$ is a factor that modulates the hardness of the interacting pair~\cite{pirani04:37,pacifici13:2668}.
This newly introduced parameter is what makes ILJ potential able to indirectly take into account some effects of atom clustering, induction and charge transfer and to improve the Lennard-Jones function in the asymptotic region.\\
Bearing in mind that charge transfer and induction effects may be important in the interaction between H$_2$O and CO$_2$ and N$_2$ a careful separate characterization of each contribution was
performed. Charge transfer effects in the perturbative limit were taken into account indirectly by lowering the value of $\beta$ as discussed, for instance, in ref.~\cite{yusuf18:16195}. In addition, induction due to the permanent water dipole was estimated and incorporated explicitly using the following semiempirical asymptotic expression (in meV)

\begin{equation}
V_{ind}(r) = -2140 \sum_{i=1}^{3} \left(\frac{3\cos^2{\gamma}+1}{2R_{OW- X_{i}}}\right)\alpha_{i}\label{eq:3}
\end{equation}
\noindent
which is applicable because of the small dimension of water with respect to the related intermolecular distances. In Eq.~\ref{eq:3}, the left coefficient -2140 (that incorporates the square of the water dipole moment value) is given in meV$\cdot\si{\angstrom}^{3}$, X$_{i}$ refers each to the C, N or to the O atoms of CO$_2$ and N$_2$, $\alpha_{i}$ is the polarizability (in $\si{\angstrom}^{3}$) associated with them and $\gamma$ is the angle formed by the $R_{OW- X_{i}}$ vector and the dipole moment of H$_2$O.

In the present system, the ILJ potential and the electrostatic interactions cutoff distance was set to 15 \AA.
The electrostatic interactions were calculated using the Ewald method, present in the DL\_POLY 2 code~\cite{dlpoly:web}.
All molecular dynamics calculations were performed using the DL\_POLY 2.
The system was studied in the canonical NVT ensemble using the Nose-Hoover thermostat and periodic boundary conditions in all directions.
At first, two simulations with only CO$_{2}$ were run with using ILJ and ILJ coupled with induction. Both lasted 5.5 ns after 0.5 ns of equilibration period with a time step of 1 fs. 
It was observed that the simulation with ILJ and induction took about 10 times longer than the calculation where only the ILJ potential was involved.
Then eight shorter calculations were run in order to analyze why the simulations with ILJ and induction were many times longer than the simulations employing only the ILJ potential.
They all lasted 600 fs after 50 fs of equilibration period and a time step of 1 fs.
Half of those calculations involved the box containing only CO$_{2}$ and the other part was done in the system with the gas mixture. 
The simulations were performed before and after a modification in one particular routine.

\section{Computational results}

\begin{table}[h]
\hspace{-1cm}
\caption{Original CPU time for 600 steps}\label{tab1}
\includegraphics[scale=0.38]{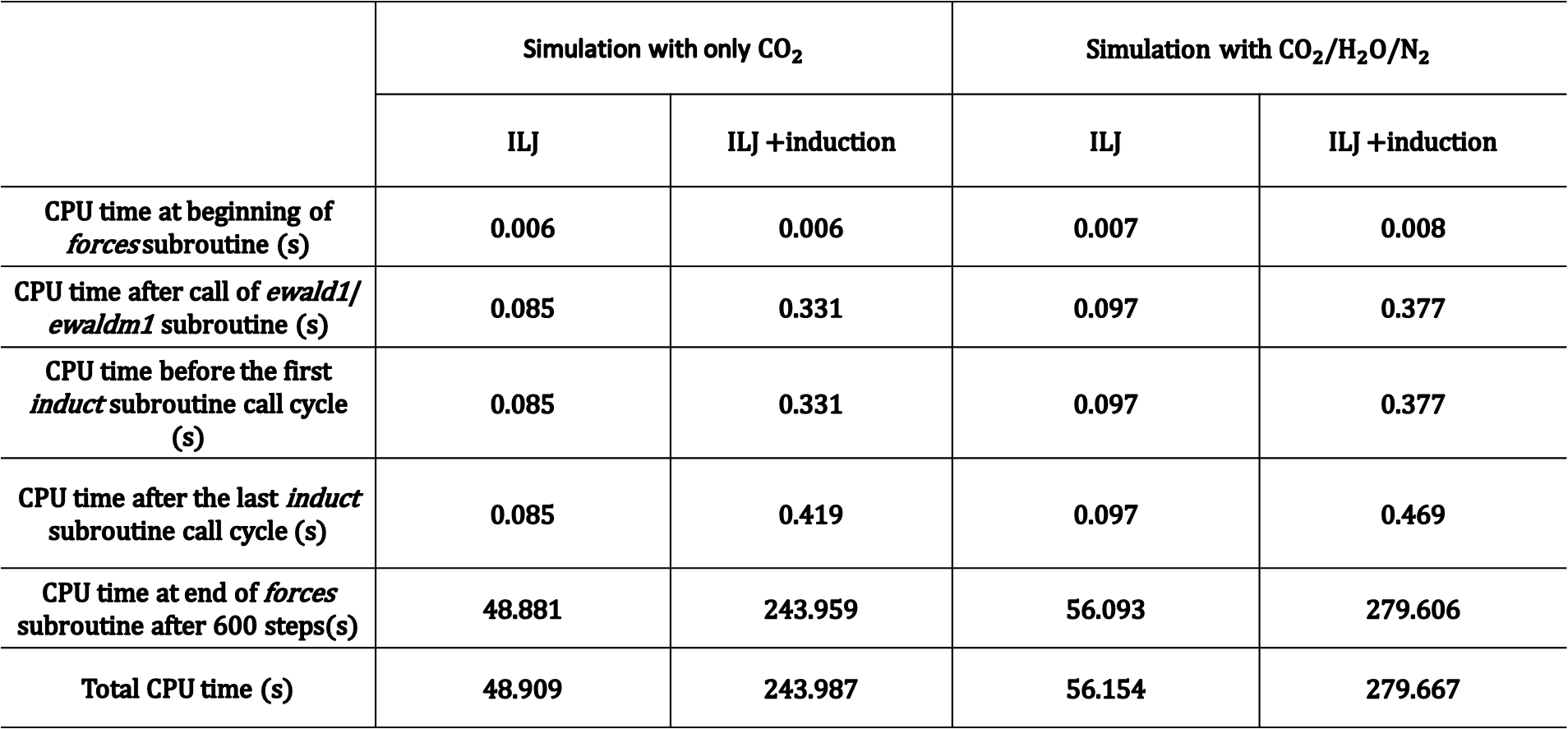}
\end{table}

As stated in the introduction, the aim of this work is to compare the time for the different simulations.
Table~\ref{tab1} reports the original simulation time before any modification was done to the code.
At first the calculations were run with the gas mixture, and it was observed that the total CPU time of the simulation with the ILJ and the induction took about five times longer than the simulation with only the ILJ potential.
Then calculations were run with only CO$_{2}$ and they were a little bit shorter compared to the ones with the gas mixture, but the simulations with ILJ+induction were still five times longer than with only ILJ.
There are no other differences in both ILJ and ILJ+induction options in the code besides the inclusion of the induction part in the latter.
This modification was done in a subroutine of the DL\_POLY2 code that calculates all the interatomic forces using the verlet neighbour list and is written in the forces\_modules.f file.
The \emph{forces} subroutine of DL\_POLY2 is a piece of code that calls other subroutines to calculate the intermolecular forces for each atom of the system. 
The subroutines subsequently called by \emph{forces} change if one is using ILJ or ILJ+induction.
If the calculation involves only the ILJ potential, the \emph{ewald1} subroutine is called.
If the simulation involves ILJ+induction, then the \emph{ewaldm1} and \emph{induct} subroutines are called.
Both \emph{ewaldm1} and \emph{induct} were written by members of the authors group in order to take the induction into account and are situated in ewald\_module.f and coulomb\_module.f respectively.

In Table~\ref{tab1}, the time is about the same for every simulation when it starts to go through \emph{forces}.
After the ewald module is called, the time of the ILJ+induction simulations is already 3.9 times higher than the time at the ILJ simulations.
Between the ewald module and the coulomb module other subroutines are called, but there is not much difference in the CPU time.
The \emph{induct} subroutine is only used in ILJ+induction simulations and because of it the CPU time becomes five longer than the ILJ simulation.
The \emph{forces} subroutine is called for each of the 600 steps of the simulation. 
The last two lines in Table~\ref{tab1} report that there is not a significant time difference between leaving the forces module at the last step and the end of the simulation.

Evidence points to something inside \emph{ewaldm1} and \emph{induct} subroutines that makes the simulation take more time.
At the moment this paper is being written, the \emph{ewaldm1} routine is still being analysed.
So discussions of the modifications in this subroutine will be left for the future.

The \emph{induct} subroutine was analysed for both the cases with CO$_{2}$ and the gas mixture.
As said before, the \emph{forces} module is called in every step of the simulation, so 600 times.
However, the \emph{induct} subroutine is not called only once in the Forces module.
In fact, \emph{induct} is run for every single atom (including pseudoatoms) in the system at every step of the simulation.
In the calculation with only CO$_{2}$, there are 909 atoms in total, so \emph{induct} is called 545400 times in the simulation.
As for the calculation with the gas mixture, the total number of atoms is 972 and the \emph{induct} subroutine is called 583200 times.
Calling that subroutine one or ten times has reported no difference in the time of the simulation, since this piece of code contains only loops and arithmetic operations.
However, when such subroutine is called hundreds of thousands of times in a simulation, it begins to heavily impact in the full CPU time.
Moreover, a further analysis of the coulomb module has shown that calling the routine for most of the atoms in the system is useless: there are if statements inside \emph{induct} that only applies to carbon and oxygen atoms of CO$_{2}$, oxygen atoms of water and nitrogen atoms of N$_{2}$.
The proposed solution for turning this task less time consuming was to modify the \emph{forces} subroutine in order to only call \emph{induct} when treating those particular atoms.

\begin{table}[ht]
\hspace{-1cm}
\caption{CPU time for 600 steps after modification in the $Forces$ module}\label{tab2}
\includegraphics[scale=0.38]{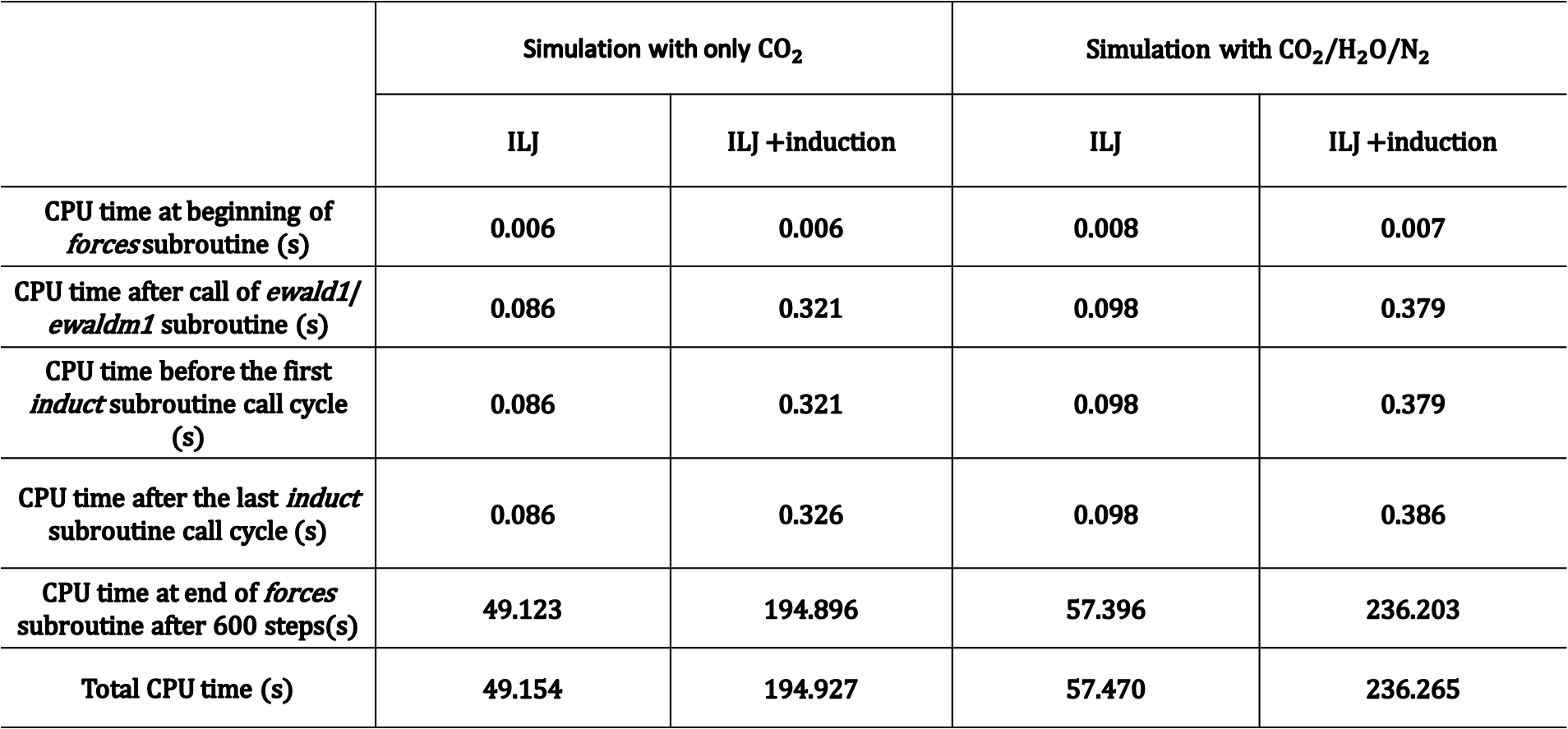}
\end{table}

Table~\ref{tab2} reports the result of this change in the code.
Calculations were run again for ILJ and ILJ+induction, with CO$_{2}$ and the gas mixture.
There were some fluctuations in the simulation times, but the most important is to look at the values of the CPU time after the last \emph{induct} call cycle.
The difference with it and the CPU time before the \emph{induct} subroutine is less than 0.01 s in both CO$_{2}$ and the gas mixture simulations.
Total CPU time for ILJ+induction is now around 4 times the Total CPU time of ILJ simulations.
When running longer calculations with ILJ+induction, for instance a simulation with 6 million steps, it is expected that the modification made on \emph{induct} will contribute to spare days of computer time.

\section{Conclusions}
A small modification in part of the code has had a considerable impact on the total simulation time.
Future modifications in the $ewaldm1$ subroutine are expected to diminish CPU time even further for calculations with ILJ and induction potentials.
While the time will not be the same as simulations with only the ILJ potential, there is at least hope that it will not be 5 or 10 times longer. 
That way, future simulations with even larger quantities of gas in the box will be possible, and the system will be described better than in simulations with only ILJ.

\section{Acknowledgements}

This project has received funding from the European Union’s Horizon
2020 research and innovation programme under the Marie Sk{\l}odowska Curie grant agreement No 811312 for the project ”Astro-Chemical Origins” (ACO).
E. V. F. A thanks the Herla Project(http://hscw.herla.unipg.it) - Universit\`{a} degli Studi di Perugia for allocated computing time. N. F.-L and A. L. thanks MIUR and the University of Perugia for the financial support of the AMIS project through the “Dipartimenti di Eccellenza” programme. N. F.-L and A. L. also acknowledges the Fondo Ricerca di Base 2017 (RICBASE2017BALUCANI) del Dipartimento di Chimica, Biologia e Biotecnologie della Università di Perugia for financial support. A. L. acknowledges financial support from MIUR PRIN 2015 (contract 2015F59J3R 002).

\bibliography{evfa_tccma_iccsa2020}{}
\bibliographystyle{splncs} 
\end{document}